\documentclass[twocolumn,preprintnumbers]{revtex4}
\usepackage{amssymb}

\usepackage[dvips]{graphicx}

\begin{document}

\title{Polarons  in 2+$\epsilon$ dimensions and giant  figure of merit of semiconducting nanolayers}

\author{A. S. Alexandrov$^{1}$ and A. M. Bratkovsky$^{2}$}
\affiliation{$^{1}$Department of Physics, Loughborough University, Loughborough LE11 3TU,
United Kingdom\\
$^{2}$Hewlett-Packard Laboratories, 1501 Page Mill Road, Palo Alto,
California 94304, United States\\
}
\begin{abstract}
Polarons - electrons  coupled with lattice vibrations
 - play a key role in the transport and optical properties of many systems of reduced dimension and dimensionality. Their confinement  affects drastically  the phonon, polaron, bipolaron and multi-polaron properties of quantum wells. Here we calculate the energy spectrum and  thermopower of  polarons confined to a potential well as a function of the well thickness. We show that the polaron mass enhancement in 2+$\epsilon$ dimensions   explains a giant thermolectric power  recently observed in doped  semiconducting nanolayers (multiple quantum wells (MQWs)), and propose a route for enhancing the performance of thermoelectric energy nanoconverters   increasing their figure of merit by more than one order of magnitude.
\end{abstract}

\pacs{71.38.-k, 72.10.Di, 72.20.Pa, 85.80.Fi}

\maketitle

While the form of the electron-phonon (Fr\"{o}hlich) Hamiltonian \cite%
{frohlich} is the same in all dimensions, the polaron mass, binding energy
and different response functions depend on the dimensionality for large
polarons \cite{PRB33-3926} and on the dimensionality and crystal lattice
structure for small polarons \cite{hague}, especially in the intermediate
coupling regime. In particular, directly following \cite{Feynman}, the
variational large polaron energy was calculated in Ref. \cite{PRB33-3926} in
different dimensions providing scaling relations valid to the second order
in the Fr\"{o}hlich coupling constant $\alpha $. In the important case of a
two-dimensional confinement (2D), scaling takes the form \cite{PRB33-3926}:
\begin{eqnarray}
E_{2\mathrm{D},p}\left( \alpha\right) &=&\frac{2}{3}E_{3\mathrm{D},p}\left( \frac{%
3\pi}{4}\alpha\right) ,
\label{scaling}\\
\frac{m_{2\mathrm{D}}^{\ast}\left( \alpha\right) }{ m _{2
\mathrm{D}}}&=&\frac{m_{3\mathrm{D}}^{\ast}\left( \frac{3\pi}{4}\alpha\right)
}{m_{3\mathrm{D}}}, \\
\mu_{2\mathrm{D}}\left( \alpha\right) &=&\mu_{3\mathrm{D}}\left( \frac{3\pi}{4}%
\alpha\right) ,
\end{eqnarray}
for the ground state binding energy (polaron level shift, $E_{p}$), polaron
mass, $m^{\ast }$, and the mobility, $\mu $ (if the scattering is dominated
by optical phonons). Here $m$ is the band mass in a rigid lattice.

These scaling relations are instrumental for a qualitative understanding of
physical properties of semiconducting layered nanostructures with polaronic
carriers, which have a potential to provide a new route for realizing
practical thermoelectric devices with the high figure of merit, $%
Z=S^{2}\sigma /\kappa $ ($S$, $\sigma $ and $\kappa $ are the thermoelectric
Seebeck coefficient, and the electrical and thermal conductivities,
respectively). Following a theoretical prediction by Hicks and Dresselhaus
\cite{hicks} that layering can increase the figure of merit due to an
increase of the density of states of confined carriers, different MQWs have
been manufactured with an enhanced $Z$. More recently Ohta \emph{et al.}
\cite{ohta} observed a giant enhancement of $S$, in Nb:SrTiO$_{3}$/SrTiO$_{3}
$ superlattices consisting of conducting Nb-doped SrTiO$_{3}$ (STO)
nano-layers confined between insulating STO layers. With decreasing
thickness of the conducting layer, the magnitude of $S$ increases to a value
about five times that of the bulk Nb:STO, reaching almost 500 $\mu $V/K for
the one unit cell layer, which is much higher than what was anticipated
theoretically \cite{hicks}.

Doped STO has been recognized as an archetypical polaronic perovskite a long
time ago \cite{eagles}. Some earlier and recent experimental observations of
the optical conductivity in the bulk Nb-doped STO \cite{mechelen} clearly
reveal the evidence of the mid-infrared optical conductivity band provided
by the polaron mechanism as in many other oxides. The effective mass of the
charge carriers is obtained by analysing the Drude spectral weight. Defining
the mass renormalization of the charge carriers as the ratio of the total
electronic spectral weight and the Drude spectral weight, a two-fold mass
enhancement is obtained, which is attributed to the electron-phonon
interaction (EPI). The missing spectral weight is recovered according to the
sum rule \cite{devreese} in the mid-infrared optical conductivity band. This
band results from EPI, traditionally associated with the polaronic nature of
the charge carriers. The effective mass obtained from the optical spectral
weights yields an intermediate electron-phonon coupling strength, $\alpha
\simeq 3$ \cite{mechelen}. Importantly, very recently Choi \emph{et al. }%
\cite{noh} using optical spectroscopy observed a three-fold mass enhancement
in Nb:SrTiO$_{3}$/SrTiO$_{3}$ superlattices compared with the bulk Nb-doped
STO as the Nb:SrTiO$_{3}$ layer thickness decreased from eleven to one unit
cell. The authors of Ref. \cite{noh} have suggested that such an increase
should be attributed to a dimensional crossover of a  polaron from 3D to 2D,
and seem to be associated with the strongly enhanced Seebeck coefficient.
Here we propose a theory of the dimensional crossover of the large polaron
from 3D to 2D providing detailed explanation of the giant thermopower
enhancement in polaronic MQWs and predicting further possibilities for
enhancing the performance of thermoelectric MQW devices.

We consider a polaron confined to an infinite potential well of the width $d$
in $z$-direction, such as Nb:SrTiO$_{3}$ layer merged into a bulk undoped
SrTiO$_{3}$. For large (or continuous, in other words) polarons the lattice
structure is irrelevant \cite{devasa}, so that we adopt an effective band
mass approximation and the canonical Fr\"{o}hlich interaction \cite{frohlich}
of doped carriers with optical phonons in ionic insulators, and also include
the confinement potential $U(z)$:
\begin{eqnarray}
H&=&-{\frac{\hbar^2\nabla^{2}}{{2m}}}+\sum_{\mathbf{q}}\left(V_{\mathbf{q}}d_{\mathbf{%
q}}e^{i\mathbf{q}\cdot\mathbf{r}}+h.c.\right)\cr
&+&\hbar \omega\sum_{\mathbf{q}}(d^{\dagger}_{\mathbf{q}}d_{\mathbf{q}}+1/2)+ U(z).
\label{frohlichH}
\end{eqnarray}
Here $d_{\mathbf{q}}$ ($d_{\mathbf{q}}^{\dagger }$) annihilates (creates)
optical phonons with the frequency $\omega $. While EPI is moderate, one can
apply the perturbation theory with respect to $V_{\mathbf{q}}$, which works
well even in the intermediate coupling regime $\alpha \lesssim 2$ as
verified by the Feynman all-coupling path-integral theory \cite{devasa}. The
quantum states of a noninteracting electron and phonons are classified with
respectively the in-plane 2D electron wave-vector, $\mathbf{k_{\parallel }}$%
, and a subband index $n$, and the 3D phonon wave-vector $\mathbf{q}$. The
unperturbed state is the vacuum $|0\rangle $ of phonons and the electron
plane wave
$|{\bf k_\parallel},n,0\rangle = {1\over{L}}e^{i{\bf k_\parallel }\cdot{\bf
\overrightarrow{\rho}}}|n\rangle|0\rangle,$
where ${\bf
\overrightarrow{\rho}}$ is the in-plane electron coordinate, and $|n\rangle$ depends on $z$. The bare energy spectrum of the electron in
a layer of the area $L^2$ is
$E_{n{\bf k_\parallel}}={\hbar^2 k_\parallel^{2}\over{2m}}+\epsilon_n$,
and $n=1,2,...$, corresponds to quantized subbands due to confinement in $z$
direction, perpendicular to the layer. EPI couples the state 
) with the energy $E_{n\mathbf{k_{\parallel }}}$ and states of a single
phonon with the momentum $\mathbf{q}$ and the electron with the energy $%
E_{n^{\prime }\mathbf{k_{\parallel }-q_{\parallel }}}$. The corresponding
matrix element is
\begin{equation}
\langle {\bf k_\parallel-q_\parallel},n';1_{\bf q}|V_{\bf q}|{\bf
k_\parallel},n;0\rangle={2\hbar\omega(\pi
\alpha)^{1/2} \langle n'|e^{i q_z
z}|n\rangle \over{q \Omega^{1/2} (2m\omega/\hbar)^{1/4}}},
\end{equation}
where
$\Omega$ is the volume of the whole sample including the doped layer
and the host bulk crystal. The renormalized second-order
 energy $\tilde{E}_{n{\bf k_\parallel}}$ is found as
\begin{eqnarray}
&&\tilde{E}_{n{\bf
k_\parallel}}=E_n({\bf k_\parallel})-{4\pi \alpha \hbar^2
\omega^2\over{(2\pi)^3 (2m \omega/\hbar)^{1/2}}} \sum_{n'}\int d^3q \cr &\times& 
 {|\langle n'|e^{i q_z z}|n\rangle|^2\over{(q_\parallel^2
+q_z^2)[\hbar^2 k_\parallel q_\parallel \cos \phi/m+\hbar^2 q^2_\parallel/2m
+\epsilon_{n'}-\epsilon_n +\hbar\omega]}}. \label{energy} \nonumber\\
\end{eqnarray}

 We consider temperatures $k_{B}T$ small compared with the characteristic
phonon energy $\hbar \omega $ and low densities of carries. In this case
there is no imaginary part of $\tilde{E}_{n\mathbf{k_{\parallel }}}$ for
sufficiently low energy subbands $n$, which means that the momentum is
conserved. Expanding over $k_{\parallel }$ and replacing the sum over $%
n^{\prime }$ for an integral, one obtains $\tilde{E}_{\mathbf{k}}=-\alpha
\hbar \omega +\hbar ^{2}k^{2}/2m_{3D}$ in the 3D system, $d\rightarrow
\infty $, with the familiar 3D polaron mass \cite{frohlich}, $m_{3D}^{\ast
}=m/(1-\alpha /6)$. Keeping only terms with $n=n^{\prime }=1$ and the matrix
element $\langle n|e^{iq_{z}z}|n\rangle =1$ in Eq.(\ref{energy}), we find $%
\tilde{E}_{\mathbf{k_{\parallel }}}=-\pi \alpha \hbar \omega /2+\hbar
^{2}k_{\parallel }^{2}/2m_{2D}$ with the polaron mass $m_{2D}^{\ast
}=m/(1-\alpha \pi /8)$ for a 2D layer, $d\rightarrow 0$. Both the polaron
binding energy and the mass satisfy the scaling relations Eqs.(\ref{scaling}%
).
\begin{figure}[tbp]
\begin{center}
\includegraphics[angle=-0, width=0.45\textwidth]
{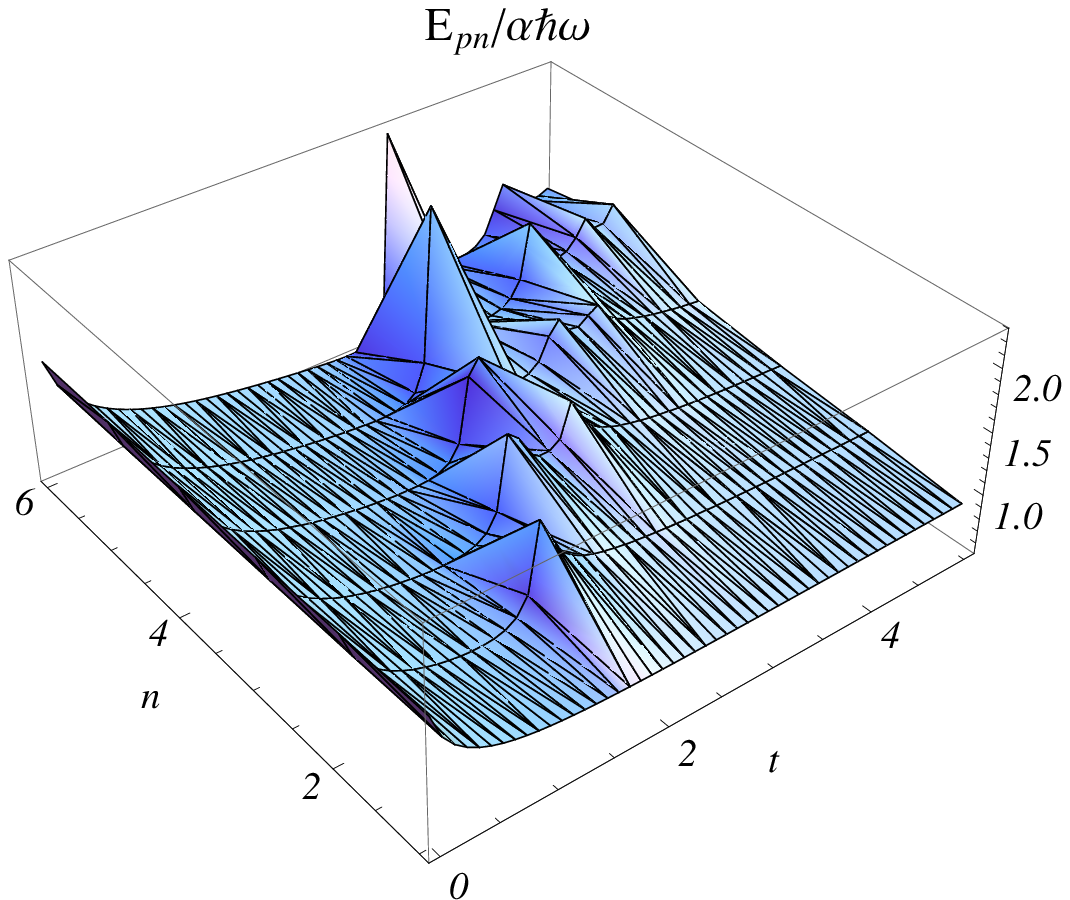}
\end{center}
\vspace{0.5cm}
\begin{center}
\includegraphics[angle=-0, width=0.4\textwidth]
{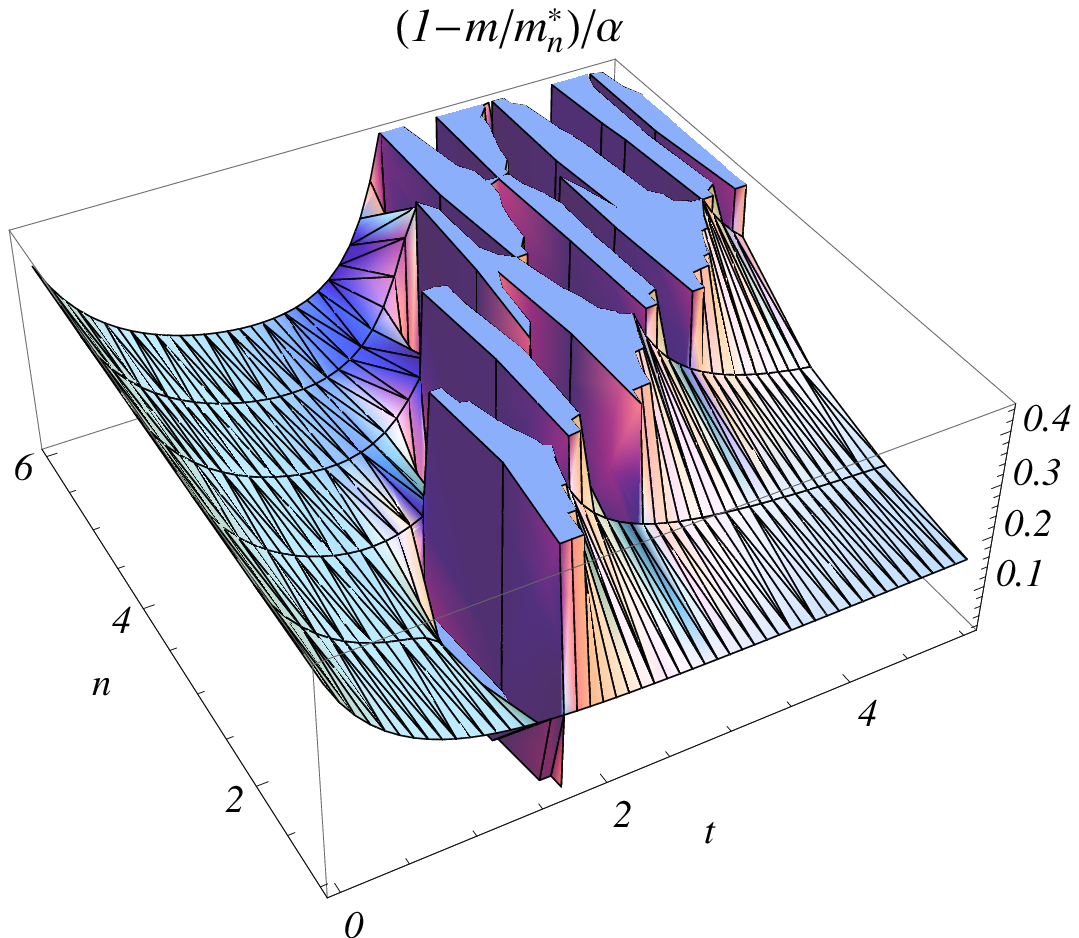}
\end{center}
\caption{Normalized polaron level shift $E_{pn}/\alpha \hbar \omega$  (top panel) and the relative  polaron mass
 renormalization $(1-m/m^*_n)/\alpha$ (bottom panel) for a number of subbands $n$ and different nanolayer thickness $t$  in the crossover region.
A sudden increase in the shift and in the mass for  sufficiently thick  nanolayers ( $t>2$) and high energy subbands is due to a breakdown of the perturbation expansion.} \label{fig1}
\end{figure}

In the crossover region, where neither 3D nor 2D limits are applied, the spectrum is  quantized into   subbands,
 \begin{equation}
\tilde{E}_{n{\bf
k_\parallel}}={\hbar^2 k_\parallel^{2}\over{2m^*_n}}+\epsilon_n-E_{pn}
\end{equation}
with the subband polaron energy shift $-E_{pn}$ and the  subband in-plane
polaron mass, $m^*_n$, found as
\begin{equation}
{E_{pn}\over{\alpha \hbar \omega}}={ t\over{\pi}} \sum_{n'}\int_0^\infty dx
\int_{-\infty} ^\infty dy
 {x|\langle n'|e^{i yu}|n\rangle|^2\over{(x^2+y^2)(x^2+n'^2-n^2+t^2)}},
\label{shift2}
\end{equation}
and
\begin{eqnarray}
{m\over{m^*_n}}&=&1-{2\alpha t^3\over{\pi}}\sum_{n'}\int_0^\infty dx
\int_{-\infty} ^\infty dy \times \cr
&& {x^3
|\langle n'|e^{i yu}|n\rangle|^2\over{(x^2+y^2)(x^2+n'^2-n^2+t^2)^3}}, \label{mass2}
\end{eqnarray}
respectively. Here, we use dimensionless phonon momenta $y=q_{z}d/\pi $, $%
x=q_{\parallel }d/\pi $, $z$-coordinate $u=\pi z/d$ of the polaron confined
to the layer of the thickness $d$, and dimensionless thickness of the layer,
$t=q_{p}d/\pi $. When polarons are confined to the infinite potential well
within $0<z<d$, the bare eigenfunctions and bare eigenvalues are $|n\rangle
=(2/d)^{1/2}\sin (\pi nz/d)$ and $\epsilon _{n}=\hbar ^{2}\pi
^{2}n^{2}/(2md^{2})$, respectively. Then, integrating first over the
transverse momentum $y$ to avoid complicated oscillating functions in Eqs.(%
\ref{shift2},\ref{mass2}) yields
\begin{equation}
{E_{pn}\over{\alpha \hbar\omega}}={ t\over{\pi^2}} \sum_{n'=1}^{\infty}
\int_{0} ^\infty dx {
 I_{n'n}(x)\over{x^2+n'^2-n^2+t^2}},
\label{shift3}
\end{equation}
and
\begin{equation}
{m\over{m^*_n}}=1-{2\alpha t^3\over{\pi^2}} \sum_{n'=1}^{\infty}
\int_{0} ^\infty dx {x^2
 I_{n'n}(x)\over{(x^2+n'^2-n^2+t^2)^3}}, \label{mass3}
\end{equation}
with
\begin{eqnarray}
I_{nn}(x)&=& {2(\pi x+e^{-\pi
x}-1)\over{x^2}}+ {\pi x\over{x^2+4n^2}}\cr
&+&{2(8n^2+x^2)(1-e^{-\pi
x})\over{(x^2+4n^2)^2}} \label{diagonal}
\end{eqnarray}
for the diagonal contribution, and
\begin{eqnarray}
I_{n'n}(x)&=&
{\pi x\over{x^2+k^2}}+
{\pi
x\over{x^2+m^2}}\cr
&-&{2x^2 (m^2-k^2)^2(1-(-1)^m e^{-\pi
x})\over{(x^2+m^2)^2(x^2+k^2)^2}} \label{offdiagonal}
\end{eqnarray}
for the off-diagonal terms, $n^{\prime }\neq n$, where $m=n^{\prime }+n$ and
$k=n^{\prime }-n$ ($n^{\prime },n=1,2,...\infty $). Two last terms in the
diagonal contribution, Eq.(\ref{diagonal}), and the last term in the
off-diagonal contribution, Eq.(\ref{offdiagonal}) are numerically small
about a few percent of the remaining terms. Integration over $x$ and
summation over $n^{\prime }$, performed using any standard software, yield
the shift and the mass interpolating well between 2D and 3D limits. However,
the polaron level shift and the polaron mass are different for different
subbands in $2+\epsilon $ crossover region as shown in Fig.\ref{fig1}. As
can be seen in Fig.\ref{fig1}, one encounters the familiar flattening of the
polaron dispersion and the breakdown of the perturbation theory at energies
about the phonon energy \cite{rashba} for high-energy subbands.

\begin{figure}[tbp]
\begin{center}
\includegraphics[angle=-0, width=0.45\textwidth]
{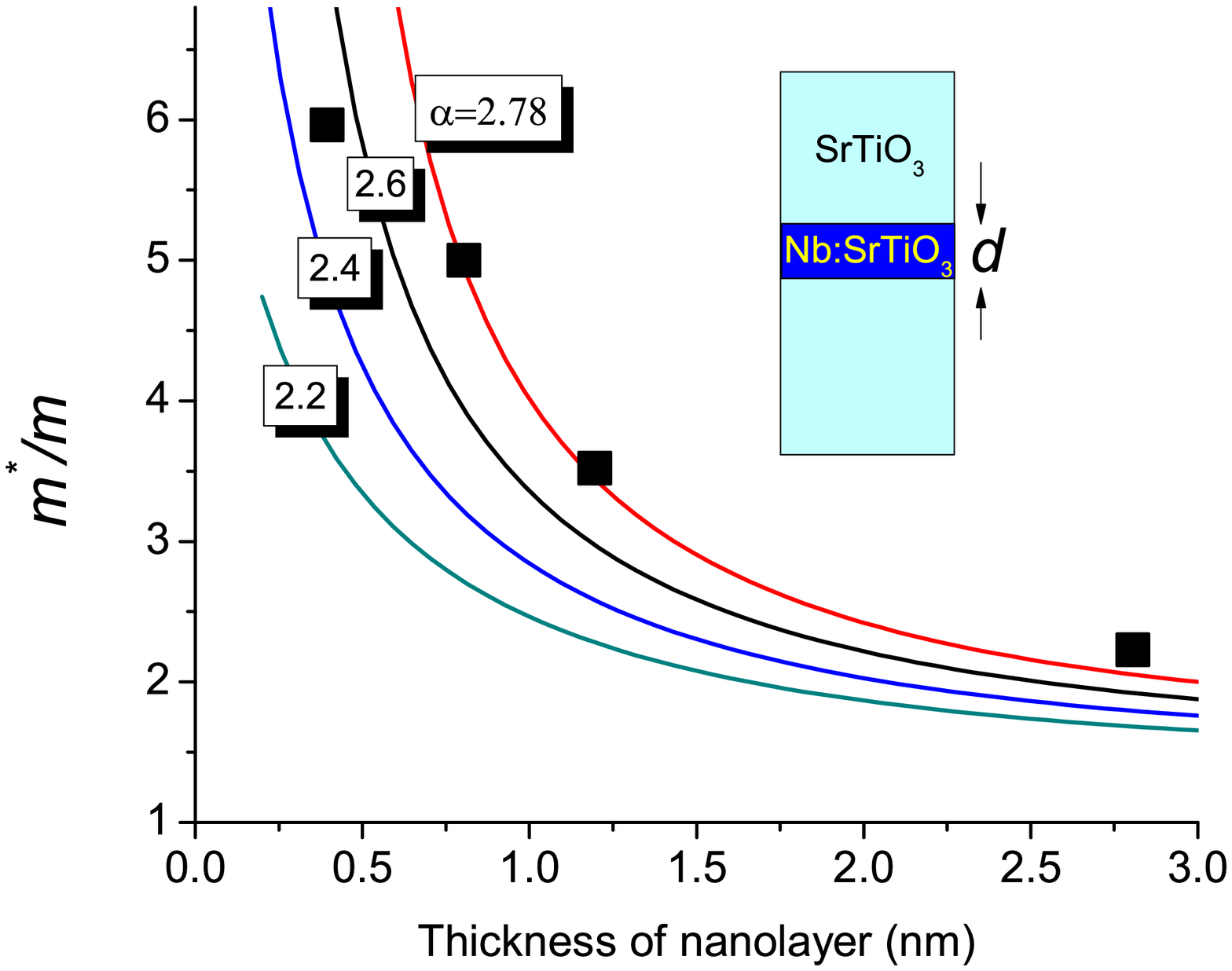}
\includegraphics[angle=-0, width=0.45\textwidth]
{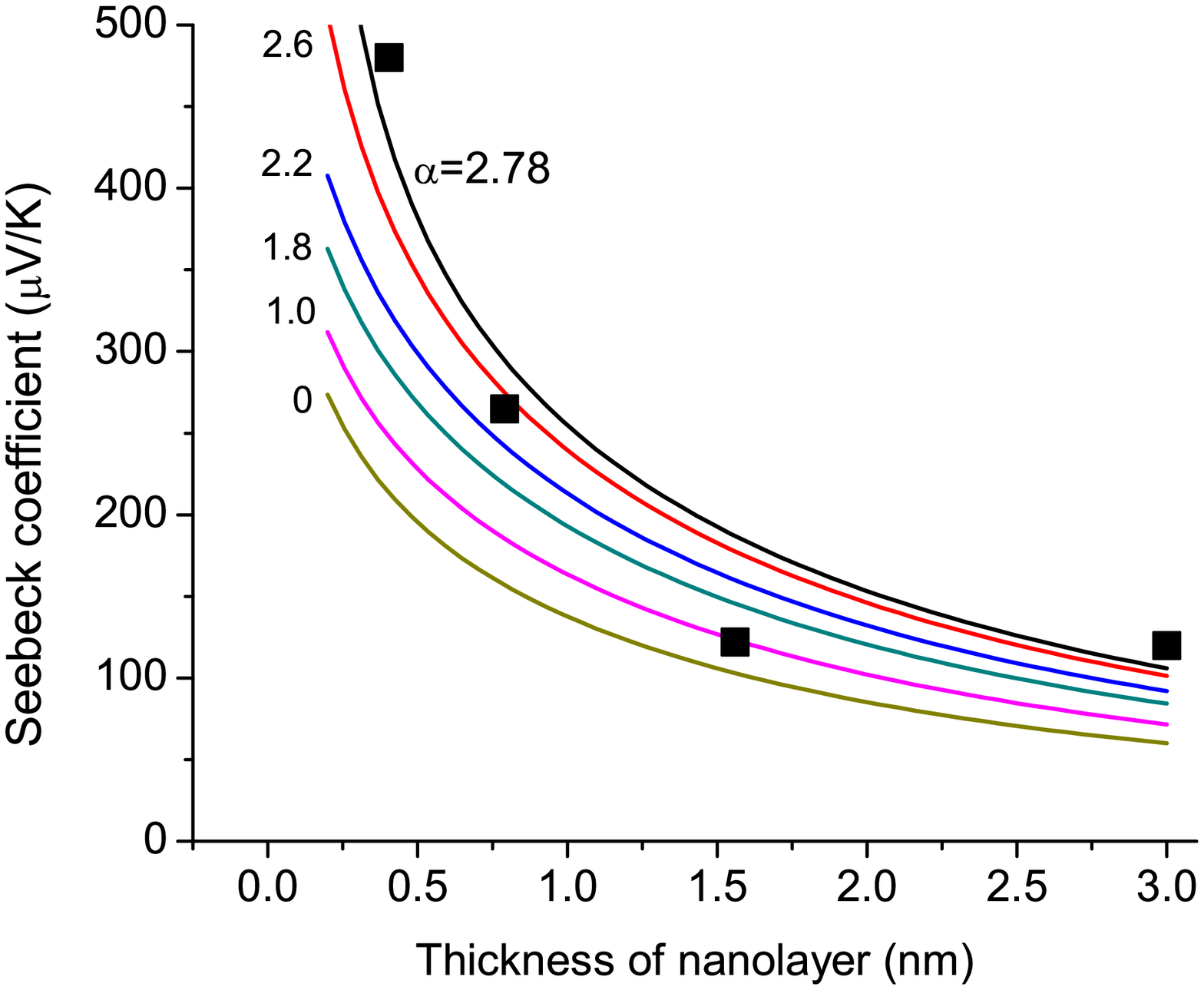}
\end{center}
\caption{ (a) Polaron mass $m^*/m$ of the lowest subband (top panel) and Seebeck coefficient $S$ (bottom panel)   in the crossover region for a few coupling constants $\alpha$  and $\hbar\omega=0.05 $ meV, $T=300$K, $m=m_e$ and $n_p=1.66 \times 10^{20}$ cm$^{-3}$ (solid lines) as the functions of the nanolayer thickness $d$ (inset in the top  panel). Symbols correspond to the experimental mass measured optically in Ref.\cite{noh} and to the experimental  Seebeck coefficient measured in Ref.\cite{ohta}.} \label{fig2}
\end{figure}

Importantly, the polaron mass changes from its 2D value to the 3D value for $%
t$ in the range of about unity, Fig.\ref{fig1}. The experimental values of the
effective thickness $t$ are about $t=0.16N$, where $N$ is the number of
atomic layers in the well \cite{ohta,noh}, if we take $\hbar \omega =0.05$
eV, $m=m_{e}$, and the lattice constant $a=0.4$ nm, which are typical of
STO. Hence, the experimental conditions are just right for a significant
change of the polaron mass with the number of layers in Nb:SrTiO$_{3}$/SrTiO$%
_{3}$ superlattices where $N$ changes from 1 to 11 \cite{ohta,noh}. The
three-fold mass enhancement with decreasing thickness \cite{noh} corresponds
to $\alpha \approx 2.8$, Fig.\ref{fig2}, which is close to optically measured value
in the bulk Nb:SrTiO \cite{mechelen}.

Equations (\ref{shift3}) and (\ref{mass3}) and the Boltzmann theory allow us
to analyze the dependence of different kinetic and thermodynamic properties
on the confinement. In particular, applying the energy-independent
relaxation time ($\tau $) approximation, one obtains the Seebeck coefficient
as:
\begin{equation}
S={k_B\over{e}} {\sum_{n,\bf k_\parallel} (\tilde{E}_{n{\bf
k_\parallel}} -\zeta) (\partial
\tilde{E}_{n{\bf
k_\parallel}}/\partial k_{\parallel})^2 \partial
f(\tilde{E}_{n{\bf
k_\parallel}})/\partial \tilde{E}_{n{\bf
k_\parallel}}\over{k_BT \sum_{n,\bf k_\parallel}
 (\partial \tilde{E}_{n{\bf
k_\parallel}}/\partial
k_{\parallel})^2 \partial f(\tilde{E}_{n{\bf
k_\parallel}})/\partial
\tilde{E}_{n{\bf
k_\parallel}}}}, \label{S}
\end{equation}
where $f(E)=1/[1+\exp (E-\zeta )/k_{B}T]$ is the Fermi-Dirac distribution
function. In the bulk system with $m_{3D}^{\ast }=2m_{e}$, the Fermi energy
is about $\zeta /k_{B}=1250$K for the polaron density $n_{p}=10^{21}$ cm$%
^{-3}$, so that carriers are almost degenerate at room temperature. However,
the system evolves from a degenerate to a classical heavier polaron gas with
decreasing nanolayer thickness because of the strong mass enhancement in the
crossover region, Fig.3. Hence one has to use the exact Fermi-Dirac
statistics in this region. Integrating over momentum in Eq.(\ref{S}) yields:
\begin{equation}
{eS\over{k_B}}= {\sum_n \left[ {\pi^2\over{3}}+(\ln
y_n)^2+2Li_2(-{1\over{y_n}}) -\ln(y_n)
\ln(1+y_n)\right]\over{\sum_n \ln(1+y_n)}}, \label{S2}
\end{equation}
where $y_n=\exp\left[(\zeta-{\pi^2 \hbar^2 n^2\over{2 m d^2}} +E_{pn})/k_B T \right]$
satisfies the following sum rule
\begin{equation}
\sum_{n=1}^{\infty} m^*_n \ln(1+y_n)={\pi \hbar^2 n_p d\over{k_B
T}}, \label{sumrule}
\end{equation}
and $Li_2(z)=\sum_{k=1}^{\infty} z^k/k^2$ is the dilogarithm
function.
 Since $\hbar \omega /k_{B}T>1$ at room temperature for typical optical
phonons in oxide ionic lattices, one can safely neglect all higher subband
contributions in Eqs.(\ref{S2}, \ref{sumrule}) compared with the lowest
subband term $n=1$ in the crossover region, $t<1$, avoiding the breakdown of
the perturbation expansion, Fig.\ref{fig1} . The thermopower, calculated
using this single-subband approximation, accounts rather well for the
experiment \cite{ohta} with the same values of the EPI coupling, which
accounts for the experimental polaron mass in Fig.\ref{fig2}. The carrier
density turns out significantly lower compared with the chemical density
presumably due to a partial localization of polarons by impurities.

Finally, let us discuss the confinement enhancement of the figure of merit.
In semiconducting MQWs such as Nb:SrTiO$_{3}$/SrTiO$_{3}$ superlattices, the
thermal conductivity is mainly due to phonons and largely independent on
doping and confinement. If we take $S\propto m^{\ast }$ then, using the
scaling Eqs.(\ref{scaling}), the maximum confinement enhancement of $Z$ is
estimated as $Z_{2D}/Z_{3D}=m_{3D}^{\ast }(3\pi \alpha /4)^{2}\mu _{3D}(3\pi
\alpha /4)/m_{3D}^{\ast }(\alpha )^{2}\mu _{3D}(\alpha )$, if polarons are
scattered by absorption of optical phonons. In this case, the mobility
depends on the coupling strength as \cite{kad} $\mu _{3D}\propto \left[
1-\alpha /6+{\mathcal{O}}(\alpha ^{2})\right] /\alpha $, so that the
enhancement, $Z_{2D}/Z_{3D}=(4/3\pi )(1-\alpha /6)/(1-\pi \alpha /8)$, is
only $1.3$ for $\alpha =2$. However, at temperatures well below the
characteristic optical phonon temperature $\hbar \omega /k_{B}$, the number
of thermal phonons is exponentially small, so that the polarons are mainly
scattered off neutral impurities since the impurity potential is screened by
heavy polarons at substantial doping. The scattering time off neutral
impurities increases with the polaron mass as $\tau \propto (m^{\ast })^{2}$
\cite{anselm}, so that the mobility is proportional to the mass, $\mu
\propto m^{\ast }$. In this case the confinement enhancement of the figure
of merit may be giant. For the coupling $\alpha =2$, $Z_{2D}/Z_{3D}=(1-%
\alpha /6)^{3}/(1-\pi \alpha /8)^{3}\approx 30$ is much larger than in a
free electron layered system \cite{hicks}. This prediction provides a
further route for enhancing the performance of thermoelectric energy
converters.

To conclude, we have developed the theory of polaron crossover from 3D to
2D. In the crossover region the polaron energy spectrum is quantized into
size subbands with the subband binding energy and polaron mass strongly
dependent on the size of the confinement nanolayer, when the thickness of
the layer is comparable or less than the large polaron radius, $d\leq 1/q_{p}
$. We have shown that this condition is satisfied in Nb:SrTiO$_{3}$/SrTiO$%
_{3}$ superlattices and proposed a detailed explanation of the giant
thermopower enhancement observed in these structures \cite{ohta,noh}. The
theory predicts a giant confinement enhancement of the figure of merit in
those polaronic MQWs where the scattering of polarons is dominated by
impurities.

We greatly appreciate illuminating discussions with J. T. Devreese, T. W. Noh and
J. Tempere, and the EPSRC (UK) support of this work (grant no. EP/H004483).

\end{document}